\documentclass[journal]{IEEEtran}
\usepackage{amsmath,amssymb,amsfonts}
\usepackage{bm}
\usepackage{cite}
\usepackage{graphicx}
\usepackage{algorithm}
\usepackage{algorithmic}
\usepackage{hyperref}
\usepackage{tikz}
\usepackage{pgfplots}
\pgfplotsset{compat=1.18}
\usetikzlibrary{arrows.meta}
\usetikzlibrary{plotmarks}
\usepackage[caption=false,font=footnotesize]{subfig}
\usepackage[table]{xcolor}
\usepackage{booktabs}
\usepackage{stfloats}

\newcommand{\ket}[1]{| #1 \rangle}

\title{  Warm-Start Quantum Approximate Optimization Algorithm for QAM MIMO Data Detection}
\author{Soumyadip Paul, Sourav Banerjee, Debanjan Bhowmik, and
Neel Kanth Kundu
\thanks{The work of Neel Kanth Kundu was supported in part by the DST INSPIRE Faculty Fellowship (Reg. No.: IFA22-ENG 344), ANRF Prime Minister Early Career Research Grant (ANRF/ECRG/2024/000324/ENS), and IIT Delhi's New Faculty Seed Grant.

Debanjan Bhowmik acknowledges research funding support from Science
and Engineering Research Board, India (under Project MTR/2023/000851),
Ministry of Education, India (under Project MoE-STARS/STARS–2/2023-
0257), and Fractal AI Research, Fractal Analytics (Level 7, Commerz II International Business Park, Oberoi Garden City, Goregaon, Mumbai, Maharashtra 400063, India).

This work was supported by the University of Melbourne through the establishment of an IBM Quantum Network Hub at the University.
}

\thanks{
Soumyadip Paul and Sourav Banerjee are with the Center for Applied Research in Electronics (CARE), Indian Institute of Technology Delhi, New Delhi-110016, India (e-mail: crz258451@care.iitd.ac.in, crz248112@care.iitd.ac.in)

Debanjan Bhowmik is with the Department of Electrical Engineering and Centre for Semiconductor Technologies (SemiX), Indian Institute of Technology Bombay, Mumbai, Maharashtra 400076, India (e-mail: debanjan@ee.iitb.ac.in).

 Neel Kanth Kundu is with CARE and the Bharti School of Telecommunication Technology and Management, Indian Institute of Technology Delhi, New Delhi-110016, India (e-mail: neelkanth@iitd.ac.in). He is also an honorary fellow at the Department of Electrical and Electronic Engineering, University of Melbourne, Melbourne, VIC-3010, Australia. (\textit{Corresponding Author: Neel Kanth Kundu}).

}

}
\begin{document}

\maketitle

\begin{abstract}
Data detection in large-scale multiple-input multiple-output (MIMO) systems with higher-order quadrature amplitude modulation (QAM) remains a challenging problem due to the exponential complexity of the classical maximum likelihood (ML) detector. This challenge is further amplified by Gray-coded modulation, which introduces nonlinear symbol-to-bit mappings and transforms the problem into a higher-order unconstrained binary optimization (HUBO) formulation. To address this problem, this paper presents a hybrid quantum–classical detection framework that leverages a warm-start linear-ramp Quantum Approximate Optimization Algorithm (WSLR-QAOA) for solving the resulting HUBO problem.  A structured warm-start based on a low-rank semidefinite relaxation, solved via a block coordinate descent (BCD) method, provides an efficient and high-quality initialization, while a linear ramp parameterization guides the QAOA optimization. Simulation results show that the proposed framework outperforms classical methods in terms of Symbol-Error-Rate (SER) and converges faster than standard QAOA, while achieving performance close to the optimal ML detector. Furthermore, the WSLR-QAOA algorithm is validated on actual IBM quantum hardware, where it achieves near-ML performance at low SNR and maintains competitive accuracy at higher SNR despite moderate degradation due to hardware noise. This demonstrates the practical potential of the HUBO-based WSLR-QAOA algorithm for large-scale MIMO data detection.

% This demonstrates the potential of the HUBO-based WSLR-QAOA algorithm for large-scale MIMO data detection.

\end{abstract}
\begin{IEEEkeywords}
Quantum approximate optimization algorithm, BM-BCD, warm start QAOA, MIMO detection, linear ramp.
\end{IEEEkeywords}
\section{Introduction}

The rapid evolution of wireless communication systems, particularly in the context of fifth-generation (5G) and beyond networks, has led to an increasing demand for higher data rates, ultra-reliable low-latency communication (URLLC), and improved spectral efficiency \cite{yang20196g}. Multiple-input multiple-output (MIMO) technology is a key enabler of these requirements, allowing multiple data streams to be transmitted simultaneously over the same frequency band. By leveraging spatial diversity and multiplexing gains, MIMO systems significantly enhance system capacity and reliability. However, these gains critically depend on the efficiency and accuracy of signal detection at the receiver~\cite{1266912}.

Maximum Likelihood (ML) detection achieves optimal performance by exhaustively searching over all possible transmitted symbol combinations \cite{hassibi2005sphere,verdu1989computational}. However, its computational complexity grows exponentially with the number of transmit antennas, making it impractical for large-scale systems employing high-order constellations such as $M$-QAM \cite{10.1007/978-981-95-0207-3_33}. Classical low-complexity alternatives such as Zero-Forcing (ZF) and Minimum Mean Square Error (MMSE) receivers \cite{5744897} suffer from performance degradation, while more advanced methods such as sphere decoding and semidefinite relaxation (SDR) \cite{1408197,5447068} exhibit limited scalability. Moreover, Gray-coded $M$-QAM introduces nonlinear symbol-to-bit mappings that transform the detection problem into a higher-order unconstrained binary optimization (HUBO) formulation \cite{Lin2010LowComplexity,doi:10.1049/ip-com:20050187}. This higher-order structure induces complex multi-variable interactions and results in a highly non-convex optimization landscape, making efficient large-scale detection particularly challenging \cite{zmgv-m2ql,Ising}. These limitations highlight a fundamental gap in achieving scalable and near-optimal MIMO detection under realistic system conditions.
% as this is explaine in section \ref{sys} .

Quantum computing has emerged as a promising paradigm for addressing such combinatorial optimization problems. Grover’s search-based quantum MIMO detection methods have also been investigated, offering quadratic speedup but relying on problem-specific quantum oracles that are difficult to realize in practice\cite{10838522,10044091}. Early also efforts, including quantum annealing-based approaches have demonstrated the feasibility of mapping large-scale MIMO detection onto quantum hardware\cite{9839195,10.1145/3341302.3342072,10838522,PhysRevX.7.021027}, primarily for simplified modulation schemes such as BPSK. However, extending these approaches to practical Gray-coded $M$-QAM systems introduces higher-order interactions that cannot be directly handled within quadratic formulations. In parallel, alternative non-von Neumann approaches such as classical oscillator Ising machines based methods have demonstrated the potential for near-logarithmic scaling and highly energy-efficient heuristic solutions for large-scale MIMO detection problems\cite{jadia2026symbol}. In this context, the Quantum Approximate Optimization Algorithm (QAOA) provides a flexible hybrid quantum-classical framework capable of directly addressing higher-order optimization problems without requiring QUBO reductions \cite{farhi2014quantumapproximateoptimizationalgorithm,PhysRevX.10.021067,BLEKOS20241,9803257}. Prior works have explored quantum-assisted ML detection using Grover adaptive search and HUBO-based formulations \cite{10044091}, as well as QAOA-based approaches and warm-start strategies \cite{Egger2021warmstartingquantum,10602129}.

However, most existing QAOA-based MIMO detection studies focus on simplified modulation schemes such as BPSK, where the problem structure reduces to quadratic formulations that are significantly easier to handle, like Sherrington-Kirkpatrick
Model for QAOA\cite{10494217,10829540}. These approaches do not extend naturally to practical communication systems employing Gray-coded $M$-QAM, where higher-order interactions dominate and the optimization landscape becomes substantially more complex. In addition, a significant research gap persists between theoretical developments and practical deployment on noisy intermediate-scale quantum (NISQ) devices \cite{10044091}. Existing methods often rely on idealized assumptions such as noise-free circuits, expectation-value-based evaluation, or simplified QUBO models, and they suffer from non-convex parameter landscapes, sensitivity to initialization, and limited scalability \cite{9803257}. The impact of hardware noise and finite-shot sampling is also frequently neglected. To address these challenges, this work proposes a practical, hybrid quantum-classical framework for MIMO data detection with higher-order modulation schemes, aligned with realistic hardware constraints. The main contributions of the paper are summarized as follows:
% The key innovations are summarized as follows:
\begin{itemize}
      \item We present a generalized higher-order unconstrained binary optimization formulation for the $M$-QAM MIMO detection problem and map it directly to a quantum Hamiltonian, enabling efficient representation of Gray-coded $M$-QAM symbols without oracle-based assumptions \cite{10044091}.
      
    \item We propose a hybrid classical-quantum QAOA algorithm for solving the MIMO data detection problem, which uses a problem-informed initialization based on a classical Burer--Monteiro block coordinate descent (BM-BCD) algorithm to generate soft estimates that guide the quantum state toward promising regions of the solution space. The structured warm-start mixer replaces the conventional transverse-field mixer to improve exploration of the non-convex energy landscape.
    
    \item We employ a deterministic linear ramp parameterization to eliminate costly per-instance classical optimization while improving convergence behavior \cite{Xiao2025QAOATruss,MontanezBarrera2025LinearRampQAOA,l5r4-zcqv}.
    
    \item We perform numerical simulations to compare the performance of the proposed warm start, linear ramp QAOA (WSLR-QAOA) algorithm with the standard QAOA and other classical baseline MIMO detectors.  We also presented a hardware-aware evaluation framework for the WSLR-QAOA algorithm by incorporating depolarizing noise and finite-shot sampling to ensure consistency with NISQ constraints \cite{nielsen2010quantum,bergholm2018pennylane}. Finally, we also test the performance of WSLR-QAOA algorithm on a real IBM Fez quantum processor.
\end{itemize}
Our simulation results reveal that the proposed WSLR-W-QAOA algorithm outperforms classical methods in terms of Symbol-Error-Rate (SER) and converges faster than standard QAOA, while achieving performance close to the optimal Maximum Likelihood (ML) detector. Experimental results on the IBM Fez quantum processor further demonstrate the practical viability of the proposed approach, where WSLR achieves performance close to the ML detector at low SNR, indicating strong robustness to hardware noise. As SNR increases, a moderate performance gap emerges due to increased sensitivity to quantum noise while still maintaining competitive detection accuracy.
Collectively, these contributions bridge the gap between theoretical quantum algorithm-based MIMO detection and practical implementation while extending QAOA-based MIMO detection from simplified BPSK settings to realistic $M$-QAM systems. 

The rest of the paper is organized as follows. Section~\ref{sys} presents the system model and HUBO formulation for ML MIMO data detection with $M$-QAM modulation in both classical and quantum settings. Section~\ref{wsqaoa} introduces the BM-BCD approach, the warm-start strategy, and the proposed linear ramp QAOA framework. Section~\ref{sim} presents the numerical simulation results and performance evaluation under both ideal and hardware-realistic conditions. Finally, Section~\ref{conc} presents the concluding remarks and future research directions.

{\em Notations}: Matrices and vectors are denoted by boldface uppercase  ($\boldsymbol{A}$), and lowercase letters($\boldsymbol{a}$), respectively. The transpose of a matrix $\boldsymbol{A}$ is denoted by $\boldsymbol{A}^T$. A multivariate complex Gaussian distribution is denoted by $\mathcal{CN}(\boldsymbol{\theta}, \boldsymbol{\Gamma})$ where $\boldsymbol{\theta} \in \mathbb{C}^N$ is the mean vector and $\boldsymbol{\Gamma} \in \mathbb{C}^{N \times N}$ is the covariance matrix.

% A Gaussian random variable $X$ with mean $\mu$ and variance $V$ is denoted as $X \sim \mathcal{N}\left(\mu,V \right)$.

\begin{figure*}
\centering

     \includegraphics[width=1\linewidth]{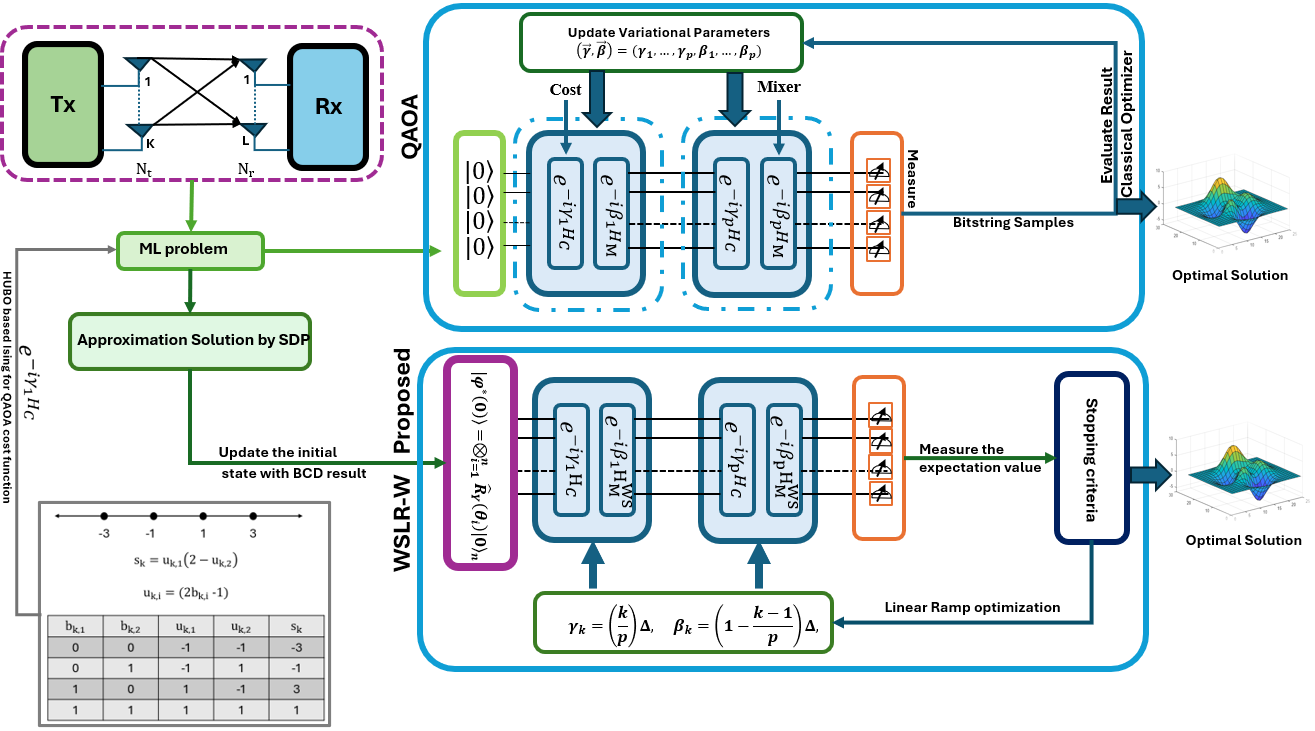}

\caption{The figure illustrates a unified quantum–classical pipeline for solving the MIMO detection problem. The transmitted–received signal model is first formulated as an ML detection problem, which is mapped to a HUBO-based cost Hamiltonian and processed using the QAOA. A classical approximation by block coordinate descent (BCD) is used to generate a problem-informed warm-start state that initializes the quantum circuit. The QAOA evolution consists of alternating applications of cost and mixer Hamiltonians, with variational parameters updated through a classical optimizer. In addition for WSLR-W, a structured linear ramp strategy is employed for parameter initialization to improve optimization efficiency. Measurement outcomes are used to evaluate expectation values, and the process iterates until a stopping criterion is satisfied, yielding the final approximate ML solution.}
\label{fig:1 flow}
\end{figure*}
% ================= Section I =================
\section{System Model and Problem Formulation}\label{sys}

Consider a narrowband MIMO system with $N_t$ transmit antennas and $N_r$ receive antennas.
The input–output relationship is given by
\begin{equation}
\mathbf{y} = \mathbf{H}\mathbf{x} + \mathbf{n},
\end{equation}
where $\mathbf{H} \in \mathbb{C}^{N_r \times N_t}$ is the MIMO channel matrix,
$\mathbf{x} \in \mathbb{C}^{N_t}$ is the transmitted symbol vector,
$\mathbf{y} \in \mathbb{C}^{N_r}$ is the received vector, and
$\mathbf{n} \sim \mathcal{CN}(\mathbf{0},N_0)$
is additive white Gaussian noise. Each entry of $\mathbf{x}$ is drawn independently from an $M$-QAM constellation set $\mathcal{S}_{M-QAM}$.
For QPSK (or $4$-QAM), the signal constellation is defined as
\begin{equation}
\mathcal{S}_{\mathrm{QPSK}} = \{ \pm 1\}+j\{ \pm 1 \},
\end{equation}
whereas the square M-QAM constellation is given by
\begin{equation}
\begin{aligned}
\mathcal{S}_{\mathrm{M\text{-}QAM}}
&= \{ \pm 1, \pm 3, \ldots, \pm (L-1) \} \\
&\quad + j\,\{ \pm 1, \pm 3, \ldots, \pm (L-1) \},\quad M = L^2. \\
\end{aligned}
\end{equation}
The objective at the receiver is to recover $\mathbf{x}$ from $\mathbf{y}$ given the channel knowledge $\mathbf{H}$. A schematic of the proposed hybrid quantum–classical framework for solving the MIMO detection problem is shown in Fig \ref{fig:1 flow}, and the details are explained in the subsections below.

\subsection{Maximum Likelihood Detection}

The maximum likelihood (ML) detector is given by
\begin{equation}\label{eq:ML}
\hat{\mathbf{x}}_{\text{ML}}
=
\arg\min_{\mathbf{x} \in \mathcal{S}^{N_t}}
\|\mathbf{y} - \mathbf{H}\mathbf{x}\|_2^2.
\end{equation}
Although ML detection is optimal, its computational complexity grows exponentially with $N_t$ as $O\left(|\mathcal{S}|^{N_t} \right)$
where $|\mathcal{S}|$ is the constellation size, making it impractical for large-scale
MIMO systems with higher-order QAM modulation schemes. Therefore, in \cite{Sidriopoulos2006}, the ML problem is relaxed as a semidefinite program (SDP), which is discussed in the subsection below.

\subsection{Semidefinite Relaxation (SDR) Approach}

\subsubsection{Real-Valued Reformulation}

Define 
$\mathbf{z} = \begin{bmatrix}\Re\{\mathbf{y}\}\\\Im\{\mathbf{y}\}\end{bmatrix},\\
 \mathbf{s} = \begin{bmatrix}\Re\{\mathbf{x}\}\\\Im\{\mathbf{x}\}\end{bmatrix}, 
\boldsymbol{\mathcal{M}} = \begin{bmatrix}\Re\{\mathbf{H}\} & -\Im\{\mathbf{H}\}\\\Im\{\mathbf{H}\} & \Re\{\mathbf{H}\}\end{bmatrix}$\\
then the detection problem becomes
\begin{equation}
\label{eq:real_ML_problem}
\min_{\mathbf{s} \in \mathbb{R}^{2N_t}} \|\mathbf{z} - \boldsymbol{\mathcal{M}}\mathbf{s}\|_2^2 \quad
\text{s.t.} \quad s_i \in \mathcal{S}, \quad i = 1,\dots,2N_t.
\end{equation}

\subsubsection{SDP Formulation}

Expanding the objective we obtain
\begin{equation}
\|\mathbf{z} - \boldsymbol{\mathcal{M}}\mathbf{s}\|_2^2 
= \mathbf{s}^T \boldsymbol{\mathcal{M}}^T \boldsymbol{\mathcal{M}} \mathbf{s} 
- 2 \mathbf{z}^T \boldsymbol{\mathcal{M}} \mathbf{s} 
+ \mathbf{z}^T \mathbf{z}.
\label{OB}
\end{equation}
Since \( \mathbf{z}^T \mathbf{z} \) is constant with respect to \( \mathbf{s} \),  eq.\eqref{eq:real_ML_problem} is equivalent to
\begin{equation}
\label{eq:ML_QP}
\min_{\mathbf{s}} \; \mathbf{s}^T \boldsymbol{\mathcal{M}}^T \boldsymbol{\mathcal{M}} \mathbf{s} 
- 2 \mathbf{z}^T \boldsymbol{\mathcal{M}} \mathbf{s} \quad
\text{s.t.} \quad s_i \in \mathcal{S}, \;\forall\, i.
\end{equation}
Since $\mathcal{S}$ is symmetric about the origin, for any $t \in \{-1,1\}$ the 
substitution $\mathbf{s} = t\tilde{\mathbf{s}}$ preserves feasibility. 
Setting $\mathbf{r} = [\,\tilde{\mathbf{s}}^T \; t\,]^T \in \mathbb{R}^{N}$, 
$N = 2N_t+1$, and using $t^2=1$, the quadratic term satisfies
\begin{equation}
t\,\tilde{\mathbf{s}}^T\boldsymbol{\mathcal{M}}^T\boldsymbol{\mathcal{M}}\,t\tilde{\mathbf{s}} 
= \tilde{\mathbf{s}}^T\boldsymbol{\mathcal{M}}^T\boldsymbol{\mathcal{M}}\tilde{\mathbf{s}}\
;.
\end{equation}
Therefore, \eqref{eq:ML_QP} reduces to
\begin{equation}
\label{eq:ML_QP_reduced}
\min_{\tilde{\mathbf{s}}} \; 
\tilde{\mathbf{s}}^T \boldsymbol{\mathcal{M}}^T \boldsymbol{\mathcal{M}} \tilde{\mathbf{s}} 
- 2 \mathbf{z}^T \boldsymbol{\mathcal{M}} \tilde{\mathbf{s}} \quad
\text{s.t.} \quad \tilde{s}_i \in \mathcal{S}, \;\forall\, i.
\end{equation}
Defining
\begin{equation}
Q = \begin{bmatrix}
\boldsymbol{\mathcal{M}}^T \boldsymbol{\mathcal{M}} & -\boldsymbol{\mathcal{M}}^T \mathbf{z} \\
-\mathbf{z}^T \boldsymbol{\mathcal{M}} & 0
\end{bmatrix},
\end{equation}
the problem lifts to
\begin{equation}
\label{eq:QP_lifted}
\min_{\mathbf{r}} \; \mathbf{r}^T Q \mathbf{r} \quad
\text{s.t.} \quad 
r_i \in \mathcal{S},\; i=1,\dots,N-1; \; 
r_N \in \{-1,1\}.
\end{equation}
Since $\mathcal{S}$ is non-convex, we relax the constraints to 
$1 \le \mathbf{r}_i^2 \le (L-1)^2$ and fix $\mathbf{r}_N = 1$. 
Introducing $\mathbf{X} = \mathbf{r}\mathbf{r}^T$, the relaxed SDP is given by
\begin{equation}\label{eq:sdp_problem}
\begin{aligned}
\min_{\mathbf{X}} \quad & \mathrm{Tr}(\mathbf{Q}\mathbf{X}) \\
\text{s.t.} \quad 
& \mathbf{X} \succeq 0, \\
& 1\leq X_{ii} \leq (L - 1)^2, \quad i = 1, \ldots, N-1,\\
& X_{NN} = 1 
\end{aligned}
\end{equation}

\subsection{QAOA formulation For MIMO Detection}
QAOA is based on the quantum adiabatic theorem and can
be seen as an extension of the Quantum Adiabatic Algorithm
(QAA). To formulate the QAOA cost function, the objective function for the ML problem has been used as in eq.\eqref{OB}.

% \textbf{\color{blue} as QAOA architecture shown in Fig \ref{fig:1 flow}}
% ref{OB}. 
% Rearranging the last term leads to the standard quadratic equation.
% \begin{equation}
% \|\mathbf{z} - \boldsymbol{\mathcal{M}}\mathbf{s}\|_2^2
% = \mathbf{s}^T \boldsymbol{\mathcal{M}}^T \boldsymbol{\mathcal{M}} \mathbf{s} - 2 \mathbf{z}^T \boldsymbol{\mathcal{M}} \mathbf{s} + \mathbf{z}^T \mathbf{z} .
% \end{equation}
Denote \[\mathbf{G} =\boldsymbol{\mathcal{M}}^T\boldsymbol{\mathcal{M}} \quad ,\mathbf{c} = \boldsymbol{\mathcal{M}}^T \mathbf{z},\] where \( \mathbf{G} \in \mathbb{R}^{2N_t \times 2N_t} \) and \( \mathbf{c} \in \mathbb{R}^{2N_t} \). Then, the objective of the optimization problem in eq.\eqref{eq:ML_QP} can be expressed as 
% Since \( \mathbf{z}^T \mathbf{z} \) is constant with respect to \( \mathbf{s} \), the objective of the optimization problem can be expressed as
\begin{equation}
f(\mathbf{s}) =\mathbf{s}^T \mathbf{G} \mathbf{s} - 2 \mathbf{c}^T\mathbf{s}.
\end{equation}
This quadratic function can be written in summation form as
\begin{equation}
H_C
=
\sum_{l=1}^{2N_t}\sum_{k=1}^{2N_t} G_{l,k}s_ls_k
-
\sum_{k=1}^{2N_t} 2c_k s_k,
\label{ob}
\end{equation}
where \( G_{l,k} \) denotes the \((k,l)\)-th elements of \(\mathbf{G}\) and \( c_k \)  denotes the $k-$th element of \(\mathbf{c}\).
% \subsection{Quantum Approximate Optimization Algorithm (QAOA) Formulation}

Instead of relying on continuous adiabatic evolution, the problem can be reformulated within the framework of  QAOA, which provides a gate-based and trotterized alternative suitable for near-term quantum devices \cite{farhi2014quantumapproximateoptimizationalgorithm}. The QAOA algorithm begins by initializing the quantum system in a uniform superposition over all computational basis states. This is achieved by applying Hadamard gates to each qubit, resulting in the state
\begin{equation}
|\psi_0\rangle = \frac{1}{\sqrt{2^n}} \sum_{x} |x\rangle.
\end{equation}
QAOA then proceeds by alternately applying two non-commuting operators, the cost Hamiltonian $H_C$, which encodes the objective function of the problem, and the mixer Hamiltonian $H_M$, which enables exploration of the solution space. The mixer Hamiltonian is typically defined as
\begin{equation}
H_M = \sum_{i} X_{i}
\end{equation}
where $X_i$ denotes the Pauli-X operator acting on the $i$-th qubit.
The parameterized QAOA state after $p$ layers is given by
\begin{equation}
|\psi(\boldsymbol{\gamma}, \boldsymbol{\beta})\rangle =
\prod_{k=1}^{p} e^{-i \beta_k H_M} e^{-i \gamma_k H_C} |\psi_0\rangle,
\end{equation}
where $\boldsymbol{\gamma} = (\gamma_1, \ldots, \gamma_p)$ and $\boldsymbol{\beta} = (\beta_1, \ldots, \beta_p)$ are variational parameters.

The objective is to find the optimal parameters that minimize the expectation value of the cost Hamiltonian, expressed as
\begin{equation}
\langle H_C \rangle =
\langle \psi(\boldsymbol{\gamma}, \boldsymbol{\beta}) | H_C | \psi(\boldsymbol{\gamma}, \boldsymbol{\beta}) \rangle.
\end{equation}
This optimization is typically performed using a classical optimizer in a hybrid quantum-classical loop. The final measurement of the optimized state yields a bitstring that approximates the optimal solution to the original problem.

\subsection{Transition to Higher-Order Binary Optimization (HUBO)}

 In standard binary expansion models (e.g., QPSK)\cite{10494217}, symbols are often assumed to be linear functions of bits, allowing the problem to be solved as a QUBO. However, modern wireless standards utilize Gray Coding to minimize the bit error rate. This mapping introduces multiplicative non-linearity, necessitating a move from QUBO to HUBO.

\subsubsection{Binary-to-Ising Mapping via Gray-Coded QAM}

For an $M$-QAM constellation, each real-valued symbol $s_k$ is represented using $W = \log_2 \sqrt{M}$ bits. Let $b_{k,i} \in \{0,1\}$ denote the $i$-th binary variable associated with symbol $s_k$. To enable a quantum formulation, these binary variables are transformed into bipolar variables defined as
\begin{equation}
u_{k,i} = 2b_{k,i} - 1 \in \{-1,+1\},
\end{equation}
which naturally correspond to the eigenvalues of Pauli-$Z$ operators, where $u_{k,i}=-Z_{k,i}$. Each real-valued symbol $s_k$ belongs to a point of $\sqrt{M}$ level pulse amplitude modulation (PAM) constellation, i.e., $s_k \in \{\pm1, \pm3, \dots, \pm(2^W-1) \}$. The real and imaginary parts of the $M$-QAM constellation points are represented using $\sqrt{M}$-PAM levels. The mapping from the bipolar variables $u_{k,i}$ to the real-valued PAM constellation points $s_k$ for Gray coding is a polynomial of order $W$ given by
\begin{equation}
\begin{aligned}
s_k = u_{k,1} \Big[
2^{W-1}-
 \sum_{i=2}^{W} 2^{W-i}
  \prod_{j=2}^{i} u_{k,j}
  \Big].
  \end{aligned}
  \label{eq:general_gray}
  \end{equation}
% This mapping defines the symbol $s_k$ as a function of the binary (or equivalently bipolar) variables, where $s_k$ is the $k$ position element in vector $\mathbf{s}$ and generates amplitude levels $\{\pm1, \pm3, \dots, \pm(2^W-1)\}$ under Gray coding.
% The mapping Eq \ref{eq:general_gray} for Gray-coded PAM levels is a non-linear polynomial of degree $W$, where real and imaginary axis on the gery map is represent with a PAM level as shown in figure \ref{fig:combinedQAM}. This mapping defines the symbol $s_k$ as a function of the binary (or equivalently bipolar) variables:
% \begin{equation}
% \begin{aligned}
% s_k = u_{k,1} \Big[
% 2^{W-1}-
%  \sum_{i=2}^{W} 2^{W-i}
%   \prod_{j=2}^{i} u_{k,j}
%   \Big].
%   \end{aligned}
%   \label{eq:general_gray}
%   \end{equation}
% which generates amplitude levels $\{\pm1, \pm3, \dots, \pm(2^W-1)\}$ under Gray coding.
For 16-QAM ($W=2$), this equation simplifies to:
\begin{equation}
s_k = u_{k,1}(2 - u_{k,2})
\label{gray16QAM}
\end{equation}
which can be written in terms of binary variables as:
\begin{equation}
s_k = (2b_{k,1} - 1)(3 - 2b_{k,2}).
\label{eq:gray_16qam2}
\end{equation}

By substituting this polynomial representation into the MIMO detection objective (\ref{ob}), the problem is transformed into an HUBO problem. Expanding this objective yields a Hamiltonian expressed entirely in terms of Pauli-$Z$ operators (see Appendix \ref{appb} for the derivation), given by
\begin{equation}
\begin{aligned}
H_C =
&\; 8 \sum_{l<k} G_{l,k} \, Z_{l,1} Z_{k,1} 
+ 4 \sum_{l<k} G_{l,k} \, Z_{l,1} Z_{k,1} Z_{k,2} \\
&+ 4 \sum_{l<k} G_{l,k} \, Z_{l,1} Z_{l,2} Z_{k,1} 
+ 2 \sum_{l<k} G_{l,k} \, Z_{l,1} Z_{l,2} Z_{k,1} Z_{k,2} \\
&+ \sum_{k=1}^{2N_t} 4c_k \, Z_{k,1} 
+ \sum_{k=1}^{2N_t} 4G_{k,k} \, Z_{k,2} + \sum_{k=1}^{2N_t} 2c_k \, Z_{k,1} Z_{k,2}.
\label{16QAM_ISING}
\end{aligned}
\end{equation}
This formulation establishes a complete mapping from binary variables to a quantum Ising Hamiltonian via Gray-coded symbol encoding, enabling implementation within the QAOA framework. Although the above derivation is only for $16$-QAM, the generalized Hamiltonian from Eq.\eqref{ob} and the Gray coded mapping in Eq.\eqref{eq:general_gray} can be used to formulate the Hamiltonian for any square $M$-QAM, like 64-QAM, 256-QAM, etc. 

% and the explanation for this is present in Appendix \ref{APPA}. 

\subsection{Implementation of QAOA}
Given $H_C$ from Eq \eqref{16QAM_ISING}, QAOA constructs a parameterized unitary operator depending on an angle $\gamma$ as
\begin{equation}
U(H_C, \gamma) = e^{-i \gamma H_C}. 
\end{equation}
Since all terms in $H_C$ are composed of Pauli-$Z$ operators, they are diagonal in the computational basis and therefore commute with each other. As a result, the unitary operator can be decomposed into a product of exponentials of individual interaction terms as given in Eq \eqref{HC} at the bottom of next page. 
\begin{figure*}[!b]
\noindent\rule{\textwidth}{0.4pt}
\begin{equation}
\begin{aligned}
U(H_C, \gamma)
&= \prod_{l<k} e^{-i \gamma \, 8 G_{l,k} \, Z_{l,1} Z_{k,1}}
 e^{-i \gamma \, 4 G_{l,k} \, Z_{l,1} Z_{k,1} Z_{k,2}}  e^{-i \gamma \, 4 G_{l,k} \, Z_{l,1} Z_{l,2} Z_{k,1}}  e^{-i \gamma \, 2 G_{l,k} \, Z_{l,1} Z_{l,2} Z_{k,1} Z_{k,2}} \\
&\times \prod_{k} e^{-i \gamma \, 4 c_k \, Z_{k,1}}  e^{-i \gamma \, 4 G_{k,k} \, Z_{k,2}} 
\, e^{-i \gamma \, 2 c_k \, Z_{k,1} Z_{k,2}}.
\end{aligned}
\label{HC}
\end{equation}
\end{figure*}
For notational simplicity, each exponential term can be represented as an elementary unitary operator acting on a subset of qubits, i.e.,
\begin{equation}
U(H_C, \gamma) = \prod_{\alpha} U(c_\alpha, \gamma), 
\end{equation}
where each $U(c_\alpha, \gamma) = e^{-i \gamma c_\alpha \prod_{i \in S_\alpha} }Z_{i}$ corresponds to a multi-qubit interaction defined over a subset of qubits $S_\alpha$ with coefficient $c_\alpha$.

Based on the initial Hamiltonian $H_M$, the mixer unitary is defined as
\begin{equation}
U(H_M, \beta) = e^{-i \beta H_M} = \prod_{k=1} e^{-i \beta X_{k}},
\label{eq:mixer}
\end{equation}
where $\beta \in [0, \pi]$. The parameters $\gamma$ and $\beta$ represent the evolution angles of the quantum system and are used to construct the parameterized QAOA circuit. Following the variational quantum eigensolver (VQE) principle, QAOA prepares a parameterized quantum state using alternating applications of the cost and mixer unitaries. For a circuit depth $p$, the resulting quantum state is given by
\begin{equation}
\begin{aligned}
|\psi_p(\boldsymbol{\gamma}, \boldsymbol{\beta})\rangle =
U(H_M, \beta_p) U(H_C, \gamma_p)\cdots \\
\cdots
U(H_M, \beta_1) U(H_C, \gamma_1)
|\psi(0)\rangle,
\end{aligned}
\end{equation}
where $\boldsymbol{\gamma} = [\gamma_1, \dots, \gamma_p]$ and $\boldsymbol{\beta} = [\beta_1, \dots, \beta_p]$.

% ================= Section III =================

% \begin{table}[t]
% \centering
% \caption{Computational Complexity Comparison}
% \begin{tabular}{|c|c|}
% \hline
% Method & Time Complexity \\ \hline
% SDP (Interior-Point) & $O(N^6 \log(1/\varepsilon))$ \\ \hline
% ADMM--SDP & $O(T_{\mathrm{ADMM}}\,N^3)$ \\ \hline
% BM--BCD & $O(T_{\mathrm{BCD}}\,N^2 K)$ \\ \hline
% \end{tabular}
% \end{table}

\section{Warm Start QAOA Algorithm} \label{wsqaoa}
Standard QAOA initializes the quantum state in the uniform superposition 
\(
|+\rangle^{\otimes n}
\) where $n$ is number of qubit and $n=N_t\log_2{M}$,
which assigns equal probability to all possible binary configurations. 
However, in detection problems such as MIMO symbol detection, a good classical 
approximation to the optimal solution is often available. Warm-start QAOA 
leverages this information to bias the initial quantum state toward promising 
regions of the solution space. The proposed WSLR-W QAOA algorithm is initialized using a warm-start solution obtained by solving the SDR problem in \eqref{eq:sdp_problem}. 

% The overall framework of the proposed QAOA-based MIMO data detection scheme is illustrated in Fig.~\ref{fig:1 flow}.

% In order to improve exploration of the non-
% convex energy landscape, the proposed QAOA algorithm is initialized using a warm start solution obtained by solving the SDR problem given by \eqref{eq:sdp_problem}. 

\subsection{Burer--Monteiro block coordinate descent (BM--BCD) method} \label{bcd}
The SDR of the MIMO detection problem in
\eqref{eq:sdp_problem} results in a convex semidefinite program (SDP) with a matrix optimization variable $\mathbf{X} \in \mathbb{R}^{N \times N}$.
Although interior-point methods can solve this problem optimally, their
computational complexity grows cubically with the problem dimension,
which makes them unsuitable for large-scale MIMO systems.
As a result, first-order and low-rank reformulation-based algorithms are
commonly employed. Motivated by \cite{erdogdu2019convergencerateblockcoordinatemaximization}, we adopt   BM-BCD approach to solve \eqref{eq:sdp_problem}. 

\subsubsection{Burer--Monteiro Factorization}

Since $\mathbf{X} \succeq \mathbf{0}$, it can be factorized as
\begin{equation}
\mathbf{X} = \mathbf{R}\mathbf{R}^T,
\quad \mathbf{R} \in \mathbb{R}^{N \times K},
\end{equation}
where $K \ll N$.
% \textbf{\color{blue} (It seems that $K=1$ from the line above Eq.(17)?)}{\color{red}We are solving the relaxed SDP; therefore unit rank constraint does not hold, $rank(X) = K$}.
Let $\mathbf{r}_i^T$ denote the $i$-th row of $\mathbf{R}$.
Then,
\begin{equation}
X_{ii} = \|\mathbf{r}_i\|^2.
\end{equation}
Substituting this factorization into \eqref{eq:sdp_problem}, we obtain
\begin{equation}
\label{eq:bm_problem}
\begin{aligned}
\min_{\mathbf{R} \in \mathbb{R}^{N \times K}} \quad
& \mathrm{Tr}(\mathbf{Q}\mathbf{R}\mathbf{R}^T) \\
\text{s.t.} \quad
& 1 \le \|\mathbf{r}_i\|^2 \le L^2, \quad i = 1,\ldots,N-1, \\
& \|\mathbf{r}_N\|^2 = 1 .
\end{aligned}
\end{equation}

\subsubsection{Block Coordinate Descent}

Using the trace identity, the objective function can be written as
\begin{equation}
\mathrm{Tr}(\mathbf{Q}\mathbf{R}\mathbf{R}^T)
=
\sum_{i=1}^{N} \sum_{j=1}^{N}
q_{ij}\, \mathbf{r}_i^T \mathbf{r}_j .
\end{equation}
When all rows except $\mathbf{r}_i$ are fixed, the objective as a function
of $\mathbf{r}_i$ reduces to
\begin{equation}
\label{eq:subproblem}
f(\mathbf{r}_i)
=
q_{ii}\|\mathbf{r}_i\|^2
+
2 \mathbf{r}_i^T
\sum_{j \neq i} q_{ij}\mathbf{r}_j
+
\text{const}.
\end{equation}
The unconstrained minimizer of \eqref{eq:subproblem} is given by
\begin{equation}
\label{eq:unconstrained_update}
\tilde{\mathbf{r}}_i
=
- \frac{1}{q_{ii}}
\sum_{j \neq i} q_{ij}\mathbf{r}_j .
\end{equation}

\subsubsection{Projection Step}

The unconstrained solution is projected onto the feasible set.
For $i = 1,\ldots,N-1$, we apply
\begin{equation}
\mathbf{r}_i^{(t+1)}
=
\Pi_{\{1 \le \|\mathbf{r}\| \le L\}}
\left( \tilde{\mathbf{r}}_i \right),
\end{equation}
where $\Pi(\cdot)$ denotes the Euclidean projection. For the last row, the unit-norm constraint yields
\begin{equation}
\mathbf{r}_N^{(t+1)}
=
\frac{\tilde{\mathbf{r}}_N}{\|\tilde{\mathbf{r}}_N\|}.
\end{equation}
The above steps are repeated until convergence, and the final solution is
recovered as $\mathbf{X} = \mathbf{R}\mathbf{R}^T$. The entire BM-BCD algorithm is summarized in Algorithm \ref{alg:bmbcd}.

\begin{algorithm}[t]
\caption{BM--BCD for SDR Problem }
\label{alg:bmbcd}
\begin{algorithmic}[1]
\REQUIRE $\mathbf{Q} \in \mathbb{R}^{N \times N}$,  bound $L$, rank $K$.
% \textbf{\color{blue}(How to Choose the rank $K$? What does $L$ denote: Is it the M-QAM parameter in eq.(3)?)}

\ENSURE $\mathbf{X} \approx \mathbf{R}\mathbf{R}^\top$

\STATE Initialize $\mathbf{R} \in \mathbb{R}^{N \times K}$
\REPEAT
    \FOR{$i=1$ to $N$}
        \STATE $\tilde{\mathbf{r}}_i \leftarrow 
        -\frac{1}{Q_{ii}} \sum_{j \neq i} Q_{ij}\mathbf{r}_j$
        \IF{$i < N$}
            \STATE $\mathbf{r}_i \leftarrow 
            \Pi_{1 \le \|\mathbf{r}\| \le L}
            (\tilde{\mathbf{r}}_i)$
        \ELSE
            \STATE $\mathbf{r}_N \leftarrow 
            \tilde{\mathbf{r}}_N / \|\tilde{\mathbf{r}}_N\|$
        \ENDIF
    \ENDFOR
\UNTIL{convergence}
\STATE $\mathbf{X} \leftarrow \mathbf{R}\mathbf{R}^\top$ 
\end{algorithmic}
\end{algorithm}

\begin{algorithm}[!t]
\caption{WSLRW-QAOA for MIMO Detection}
\label{alg:wslrw_qaoa}
\begin{algorithmic}[1]

\REQUIRE 
$ \boldsymbol{\mathcal{M}} \in \mathbb{R}^{2N_r \times 2N_t}$, $\mathbf{z} \in \mathbb{R}^{2N_r}$, depth $p$, shots $S$, parameter set $\mathcal{S}_\Delta$

\ENSURE 
$\hat{\mathbf{r}} \in \mathcal{A}^{2N_t}$

\vspace{0.1cm}
\STATE \textbf{HUBO Construction:}
\STATE $\mathbf{G} \gets \boldsymbol{\mathcal{M}}^T \boldsymbol{\mathcal{M}}, \quad \mathbf{c} \gets \mathbf{H}^T \mathbf{z}$
\STATE $f(\mathbf{s}) = \mathbf{s}^T \mathbf{G}\mathbf{s} - 2\mathbf{c}^T \mathbf{s}$

\STATE Apply Gray mapping as in Eq \eqref{eq:general_gray}
\STATE Construct Ising Hamiltonian: $H_C$ as in Eq \eqref{16QAM_ISING}

\STATE Let $\mathcal{W}$ be the set of all scalar coefficients in $H_C$
\STATE Scale: $H_C \gets \alpha H_C,\quad \alpha = \frac{1}{\max |\mathcal{W}|}$

\vspace{0.1cm}
\STATE \textbf{Warm-Start Initialization:}
\STATE Solve $\mathbf{r}^* \gets \arg\min_{\mathbf{r}} \mathbf{r}^T \mathbf{G}\mathbf{r} - 2\mathbf{c}^T \mathbf{r}$ using Algorithm \ref{alg:bmbcd}.
% \hfill (BCD approximation)
\STATE Relax to probabilities: $x_{k,i} \in [0,1]$ using Eq.\eqref{eq:xi_general}.

\STATE Prepare the initial state $\ket{\psi_0}$ using Eq \eqref{eq:initial_ws}.

\vspace{0.1cm}
\STATE \textbf{Parameter Exploration with Linear Ramp QAOA:}
\STATE $E_{\min} \gets \infty$

\FOR{$\Delta \in \mathcal{S}_\Delta$}

\STATE Initialize parameters:$\ket{\psi} \gets \ket{\psi_0}$

\FOR{$k = 1$ to $p$}
    \STATE $\gamma_k = \frac{k}{p}\Delta,\quad \beta_k = \left(1 - \frac{k-1}{p}\right)\Delta$
    \STATE $\ket{\psi} \gets e^{-i \gamma_k H_C}\ket{\psi}$
    
    \FOR{$i = 1$ to $n$}
        \STATE Define the warm-start mixer $H_{\text{M}}^{WS}$ as in Eq \eqref{WSM}
        \STATE $\ket{\psi} \gets e^{-i \beta_k H_{\text{WS}}^{(i)}} \ket{\psi}$
    \ENDFOR
\ENDFOR

\STATE Sample $\{\mathbf{b}_s\}_{s=1}^S \sim |\langle \mathbf{b} | \psi \rangle|^2$

\STATE $\mathbf{b}^* = \arg\max_{\mathbf{b}} \; \text{count}(\mathbf{b})$

\IF{$f(\mathbf{s}(\mathbf{b}^*)) < E_{\min}$}
    \STATE $E_{\min} \gets f(\mathbf{s}(\mathbf{b}^*))$
    \STATE $\mathbf{b}_{\text{opt}} \gets \mathbf{b}^* $
\ENDIF

\ENDFOR

\vspace{0.1cm}
\STATE \textbf{Decoding}
\STATE $\hat{\mathbf{r}} \leftarrow (\mathbf{b}_{\text{opt}})$ \textbf{(Inverse Gray Mapping)}

\RETURN $\hat{\mathbf{r}}$

\end{algorithmic}
\end{algorithm}

\subsection{Warm-Start QAOA Initialization}

% Standard QAOA initializes the quantum state in the uniform superposition 
% \(
% |+\rangle^{\otimes B}
% \),
% which assigns equal probability to all possible binary configurations. 
% However, in detection problems such as MIMO symbol detection, a good classical 
% approximation to the optimal solution is often available. Warm-start QAOA 
% leverages this information to bias the initial quantum state toward promising 
% regions of the solution space.

To obtain the warm-start probability vector, the binary variables 
$b_{k,i} \in \{0,1\}$ are relaxed to continuous values $x_{k,i}^* \in [0,1]$, 
which are interpreted as the probability that bit $b_{k,i} = 1$. 
For 16-QAM with $W = 2$ bits per dimension, the real and imaginary components of each transmitted symbol $s_k$ are encoded by two bits $(b_{k,1}, b_{k,2})$ via the Gray mapping \eqref{eq:gray_16qam2}. The soft bit probabilities are obtained via relaxed  estimate $r_k^*$  from BCD with a temperature parameter 
$T > 0$ controlling the sharpness of the mapping given by
\begin{equation}
x_{k,1}^* = \frac{1}{1 + e^{-r_k^*/T}}, \qquad
x_{k,2}^* = \frac{1}{1 + e^{(|r_k^*| - 2)/T}},
\end{equation}
where $x_{k,1}^*$ encodes the sign of $r_k^*$, and $x_{k,2}^*$ encodes the magnitude. For general $M$-QAM with $W = \log_2\!\sqrt{M}$ bits, for each $s_k$ value, the sign bit $x_{k,1}^*$ is identical for all $M$-QAM orders, and for bits $i = 2, 3, \ldots, W$, the soft estimate follows a recursive nested absolute-value structure. Define the folded residual recursively as:
\begin{equation}
    d_1(r_k^*) = -r_k^*, \;
    d_i(r_k^*) = \bigl|d_{i-1}(r_k^*)\bigr| - 2^{W-i+1}, 
    \; i = 2, \ldots, W.
    \label{eq:folding}
\end{equation}
Then the general soft bit estimate for bit $i \geq 2$ is:
\begin{equation}
    x_{k,i}^* = \frac{1}{1 + \exp\!\left(\dfrac{d_i(r_k^*)}{T}\right)},
    \quad i = 2, 3, \ldots, W,
    \label{eq:xi_general}
\end{equation}
where $d_i(r_k^*)$ unfolds the Gray-coded decision boundaries, 
with the decision distance halving at each bit level.

Each soft probability $x_{k,i}^*$ parameterizes a single-qubit state
\begin{equation}
    |\psi_{k,i}\rangle = \sqrt{1 - x_{k,i}^*}\,|0\rangle 
                       + \sqrt{x_{k,i}^*}\,|1\rangle,
    \label{eq:single_qubit}
\end{equation}
replacing the uniform superposition $H|0\rangle$ used in standard QAOA. The full warm-start initial state over all $n = 2N_tW$ qubits is given by
\begin{equation}
    |\psi_0\rangle = \bigotimes_{k=1}^{2N_t}\bigotimes_{i=1}^{W}
    \Bigl(\sqrt{1-x_{k,i}^*}\,|0\rangle + \sqrt{x_{k,i}^*}\,|1\rangle\Bigr).
    \label{eq:psi0}
\end{equation}
which can be prepared by applying single-qubit $R_Y$ rotation gates. Note that applying $R_Y(\theta)$ rotation to \(|0\rangle\) yields
\begin{equation}
    R_Y(\theta)|0\rangle = \cos\!\tfrac{\theta}{2}|0\rangle 
                         + \sin\!\tfrac{\theta}{2}|1\rangle\;.
\end{equation}
The rotation angle can be obtained by equating the probability of measuring $|1\rangle\ $ to $x_{k,i}^*$ i.e.,   $\sin^2(\theta_{k,i}/2) = x_{k,i}^*$, giving
\begin{equation}
    \theta_{k,i} = 2\arcsin\!\left(\sqrt{x_{k,i}^*}\right)\;.
    \label{eq:theta}
\end{equation}
Therefore, the initial state for the warm-start QAOA is prepared as
\begin{equation}
|\psi_0\rangle =
|\psi_0\rangle = \bigotimes_{k=1}^{2N_t}\bigotimes_{i=1}^{W}
R_Y\!\left(
2\arcsin(\sqrt{x_{k,i}})
\right)
|0\rangle .
\label{eq:initial_ws}
\end{equation}

After initialization, the QAOA procedure is applied by alternating 
between the cost unitary
\begin{equation}
U(H_C,\gamma) = e^{-i\gamma H_C},
\end{equation}
and a mixer unitary. In addition to the standard transverse-field mixer defined in \eqref{eq:mixer}, a problem-informed warm-start mixer is also employed, defined as
\begin{equation}
    H_M^{WS} = \sum_{k=1}^{2K} \sum_{i=1}^{W} P_{k,i},
\end{equation}
where
\begin{equation}
    P_{k,i} =
    2\sqrt{x_{k,i}^*(1-x_{k,i}^*)}\, X_{k,i}
    +
    (1-2x_{k,i}^*)\, Z_{k,i},
\end{equation}
which in matrix form is
\begin{equation}
P_{k,i} =
\begin{bmatrix}
1 - 2x_{k,i}^* & 2\sqrt{x_{k,i}^*(1-x_{k,i}^*)} \\
2\sqrt{x_{k,i}^*(1-x_{k,i}^*)} & 2x_{k,i}^* - 1
\end{bmatrix}\;.
\label{WSM}
\end{equation}
In practice, this mixer is implemented using a single-qubit unitary evolution 
\(e^{-i\beta P_{k,i}}\) for each qubit. The warm-start mixer is constructed once from \(x_{k,i}^*\) and remains fixed throughout the QAOA evolution.

For a depth \(p\), the final QAOA state is
\begin{equation}
|\psi(\boldsymbol{\gamma},\boldsymbol{\beta})\rangle =
\prod_{k=1}^{p}
U_M(\beta_k)
U_C(\gamma_k)
|\psi_0\rangle .
\end{equation}
By initializing the circuit close to a good classical solution and employing a problem-informed mixer, warm-start QAOA improves convergence and guides the quantum state toward low-energy regions of the solution space.

\subsection{Parameter Exploration via Linear Ramp Schedule}

Finding suitable variational parameters ($\boldsymbol{\gamma}$ and $\boldsymbol{\beta}$) for QAOA is challenging due to the highly non-convex optimization landscape. Traditional optimization methods may become trapped in suboptimal local minima. To mitigate this issue, a structured parameter exploration strategy based on a linear ramp schedule is employed.

The linear ramp is inspired by adiabatic quantum computing, where the system gradually evolves from an initial Hamiltonian to the problem Hamiltonian. To emulate this behavior within a discrete QAOA circuit of depth $p$, the parameters are initialized as
\begin{equation}
\gamma_k = \left(\frac{k}{p}\right) \Delta, \quad 
\beta_k = \left(1 - \frac{k-1}{p}\right) \Delta,
\end{equation}
for each layer $k \in \{1, \dots, p\}$. The parameter $\Delta$ controls the transition rate \cite{MontanezBarrera2025LinearRampQAOA}. To enhance exploration, multiple values of $\Delta$ are considered. For each value, the corresponding QAOA circuit is executed and evaluated based on the expected cost obtained from measurement samples. The parameter set yielding the lowest observed cost is selected. This approach avoids the need for computationally expensive classical optimization while enabling effective exploration of the parameter space, making it suitable for noisy quantum simulations. The overall WSLRW-QAOA algorithm for MIMO data detection is summarized in Algorithm \ref{alg:wslrw_qaoa}.

\subsection{Computational Complexity Analysis}

In this subsection, we compare the computational complexity of solving
the SDR problem using a generic interior-point method
(as implemented in CVX) and the proposed
BM--BCD algorithm.
Let $T_{\text{CVX}}$, and $T_{\text{BCD}}$ denote the
numbers of iterations required by CVX, and BM--BCD, respectively,
to achieve an $\varepsilon$-accurate solution.

\subsubsection{SDP Solved via CVX (Interior-Point Method)}

% When solved using CVX, the SDR is handled by a primal--dual
% interior-point SDP solver such as SDPT3 or SeDuMi.
% The optimization variable is a symmetric matrix
% $\mathbf{X}\in\mathbb{R}^{N\times N}$, which contains
% $\mathcal{O}(N^2)$ scalar decision variables and is constrained to lie
% in the positive semidefinite cone.
% At each iteration, a large Newton system must be solved, whose
% computational cost scales cubically with the number of variables.
% As a result, the per-iteration complexity is $\mathcal{O}(N^6)$.
% Interior-point methods are known to converge within
% $\mathcal{O}(\sqrt{N}\log(1/\varepsilon))$ iterations.
% Therefore, the overall computational complexity of the CVX-based
% solution is on the order of $\mathcal{O}(N^{6.5})$, with a memory
% requirement of $\mathcal{O}(N^4)$, which becomes prohibitive for
% large-scale MIMO detection problems.
The SDR is solved in CVX using primal–dual interior-point SDP solvers such as SDPT3 or SeDuMi. The optimization variable is a symmetric matrix $\mathbf{X}\in\mathbb{R}^{N\times N}$ with $\mathcal{O}(N^2)$ entries constrained to be positive semidefinite. Each iteration requires solving a Newton system with $\mathcal{O}(N^6)$ complexity, while convergence occurs in $\mathcal{O}(\sqrt{N}\log(1/\varepsilon))$ iterations.
Thus, the total complexity is approximately $\mathcal{O}(N^{6.5})$ with memory requirement $\mathcal{O}(N^4)$\cite{sedumi2002}, making it impractical for large-scale MIMO detection.
% \subsubsection{ADMM-Based SDP}

% For the ADMM-based approach, the dominant computational cost per
% iteration arises from the projection onto the positive semidefinite
% cone.
% This step requires an eigenvalue decomposition of an $N\times N$
% symmetric matrix, leading to a per-iteration complexity of
% \begin{equation*}
% \mathcal{O}(N^3).
% \end{equation*}
% All remaining operations, including diagonal projections and dual
% updates, have $\mathcal{O}(N^2)$ complexity and are therefore
% negligible.
% For convex problems, ADMM is known to converge at a sublinear rate of
% $\mathcal{O}(1/k)$.
% Consequently, achieving an $\varepsilon$-accurate solution requires
% \begin{equation*}
% T_{\text{ADMM}} = \mathcal{O}(1/\varepsilon)
% \end{equation*}
% iterations, resulting in a total computational complexity of
% \begin{equation*}
% \mathcal{O}\!\left(\frac{N^3}{\varepsilon}\right),
% \end{equation*}
% with a memory requirement of $\mathcal{O}(N^2)$.

\subsubsection{BM--BCD Method}

The proposed BM--BCD algorithm employs a low-rank factorization of the
SDP variable, $\mathbf{X}=\mathbf{R}\mathbf{R}^\top$, where
$\mathbf{R}\in\mathbb{R}^{N\times K}$ and $K\ll N$.
Within each outer iteration, the algorithm sequentially updates the
rows of $\mathbf{R}$.
Updating a single row requires $\mathcal{O}(NK)$ operations, and
updating all $N$ rows in one outer iteration incurs a computational cost  of $\mathcal{O}(N^2 K)$.
% \begin{align*}
% \mathcal{O}(N^2 K).
% \end{align*}
The resulting optimization problem is nonconvex; nevertheless, BM--BCD
is guaranteed to converge to a first-order stationary point, with a
typical sublinear convergence rate of $\mathcal{O}(1/k)$.
Thus, reaching an $\varepsilon$-stationary solution requires $T_{\text{BCD}} = \mathcal{O}(1/\varepsilon)$
% \begin{equation*}
% T_{\text{BCD}} = \mathcal{O}(1/\varepsilon)
% \end{equation*}
iterations, yielding a total computational complexity of
$\mathcal{O}\!\left(\frac{N^2 K}{\varepsilon}\right)$, and a reduced memory requirement of $\mathcal{O}(NK)$.

\subsubsection{Overall Comparison}

The computational and memory complexities of these two approaches are
summarized in Table~\ref{tab:complexity}.
While the CVX-based interior-point method provides a globally optimal
solution and serves as a benchmark for small-scale problems, its
computational and memory costs scale poorly with the problem dimension.
In contrast, by exploiting a low-rank factorization with $K\ll N$, the
proposed BM--BCD method achieves substantially lower computational and
memory complexity, making it well suited for large-scale MIMO detection.

\begin{table}[t]
\caption{Computational Complexity Comparison of SDR Solvers}
\label{tab:complexity}
\centering
\resizebox{\columnwidth}{!}{
\begin{tabular}{|c|c|c|c|c|}

\hline
\textbf{Method} & \textbf{Per-Iteration} & \textbf{Iterations} &
\textbf{Total Complexity} & \textbf{Memory} \\
\hline
CVX (Interior-Point) &
$\mathcal{O}(N^6)$ &
$\mathcal{O}(\sqrt{N}\log(1/\varepsilon))$ &
$\mathcal{O}(N^{6.5})$ &
$\mathcal{O}(N^4)$ \\
\hline
% ADMM--SDP &
% $\mathcal{O}(N^3)$ &
% $\mathcal{O}(1/\varepsilon)$ &
% $\mathcal{O}(N^3/\varepsilon)$ &
% $\mathcal{O}(N^2)$ \\
% \hline
BM--BCD &
$\mathcal{O}(N^2 K)$ &
$\mathcal{O}(1/\varepsilon)$ &
$\mathcal{O}(N^2 K/\varepsilon)$ &
$\mathcal{O}(NK)$ \\
\hline
\end{tabular}}
\end{table}

\subsubsection{WSLR-W Method}
The MIMO detection problem is mapped to a HUBO Hamiltonian acting on 
$n = N_t \log_2 M$ qubits. After expansion into Pauli operators, the number of terms scales as $\mathcal{O}(N_t^2)$, which dominates the cost Hamiltonian. Each QAOA layer applies the cost and mixer unitaries, with complexities of $\mathcal{O}(N_t^2)$ and $\mathcal{O}(N_t)$, respectively. Hence, the per-layer complexity is $\mathcal{O}(N_t^2)$, and for depth $p$, the total circuit complexity is $\mathcal{O}(p\,N_t^2)$.
% \begin{equation*}
% \mathcal{O}(p\,N_t^2).
% \end{equation*}
With $S$ measurement shots, each evaluation requires a total computational cost of $\mathcal{O}(S\,p\,N_t^2)$ accounting for circuit execution, measurements, and expectation estimation.
It’s also the total computational cost per evaluation.
% \begin{equation*}
% \mathcal{O}(S\,p\,N_t^2).
% \end{equation*}
Unlike conventional QAOA, which requires iterative optimization over $2p$ parameters, the proposed method employs a linear-ramp parameterization with a small set of candidates.

\begin{figure*}[ht]
    \centering
    \includegraphics[width=\textwidth]{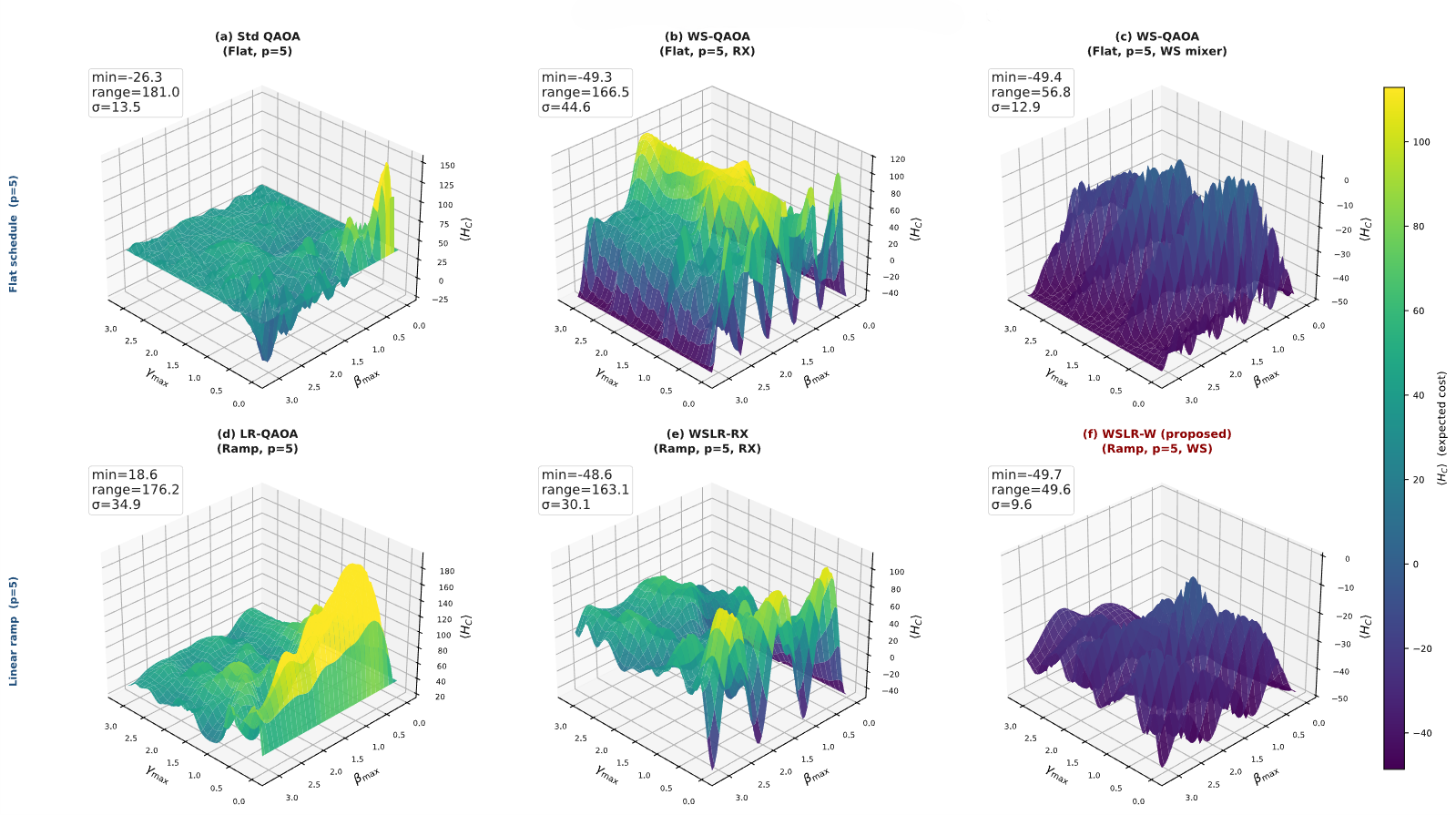}
    
    \caption{QAOA cost landscape analysis for a $2\times2$ MIMO system at
    $p=5$ layers and $\text{SNR}=7.0$~dB. \textit{Top:} Expected cost
    $\langle H_C \rangle$ evaluated over a uniform grid of parameter
    schedules $(\gamma_{\max}, \beta_{\max}) \in [0,3]^2$ for six
    algorithm variants grouped by parameter schedule (flat vs.\ linear
    ramp) and initialization/mixer strategy (standard, warm-start RX,
    warm-start WS).}
    \label{fig:landscape}
\end{figure*}

\begin{table*}[h!]
\centering
\caption{Performance comparison of different QAOA variants}
\label{perfQAOA}
\resizebox{\textwidth}{!}{
\begin{tabular}{|c|c| c| c| c| c| c| c| c| c|}
\hline
\textbf{Algorithm} & \textbf{Schedule} & \textbf{Initialization} & \textbf{Mixer} & \textbf{Min\_HC} & \textbf{Max\_HC} & \textbf{Range} & \textbf{Mean\_HC} & \textbf{Std\_Dev} \\
\hline
Std QAOA & Flat & Uniform $|+\rangle$ & RX & -26.296 & 154.734 & 181.030 & 37.533 & 13.464  \\
\hline
WS-QAOA (RX) & Flat & BCD WS & RX & -49.272 & 117.194 & 166.467 & 29.741 & 44.593 \\
\hline
WS-QAOA (WS) & Flat & BCD WS & WS & -49.424 & 7.390 & 56.815 & -34.670 & 12.896  \\
\hline
LR-QAOA & Linear Ramp & Uniform $|+\rangle$ & RX & 18.584 & 194.755 & 176.170 & 63.440 & 34.922  \\
\hline
WSLR-RX & Linear Ramp & BCD WS & RX & -48.617 & 114.465 & 163.082 & 29.604 & 30.078 \\
\hline
\textbf{WSLR-W (proposed)} & Linear Ramp & BCD WS & WS & \textbf{-49.712} & \textbf{-0.075} & \textbf{49.637} & \textbf{-33.891} & \textbf{9.632}  \\
\hline
\end{tabular}}
\end{table*}

\begin{figure*}[htbp]
    \centering
    
    % --- First Subfigure ---
    \subfloat[ 2$\times$2 
    MIMO ]{\label{fig:ser_all}
      
        \includegraphics[width=0.5\linewidth]{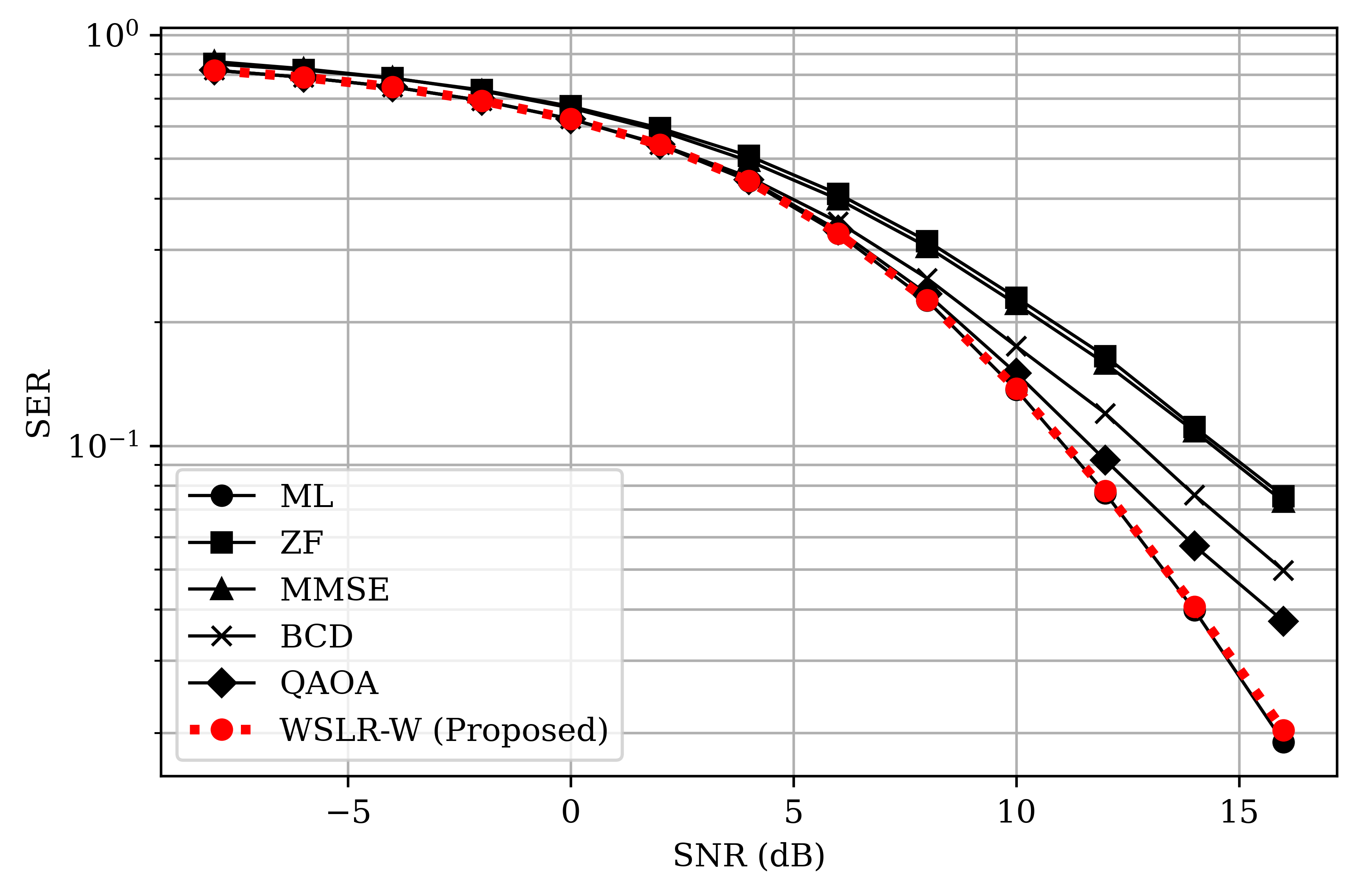}

    }
    \subfloat[ $4 \times 4$ MIMO \label{fig:mimo_qaoa_4x4}]{\label{fig:mimo_qaoa_4x4}

         \includegraphics[width=0.5\linewidth]{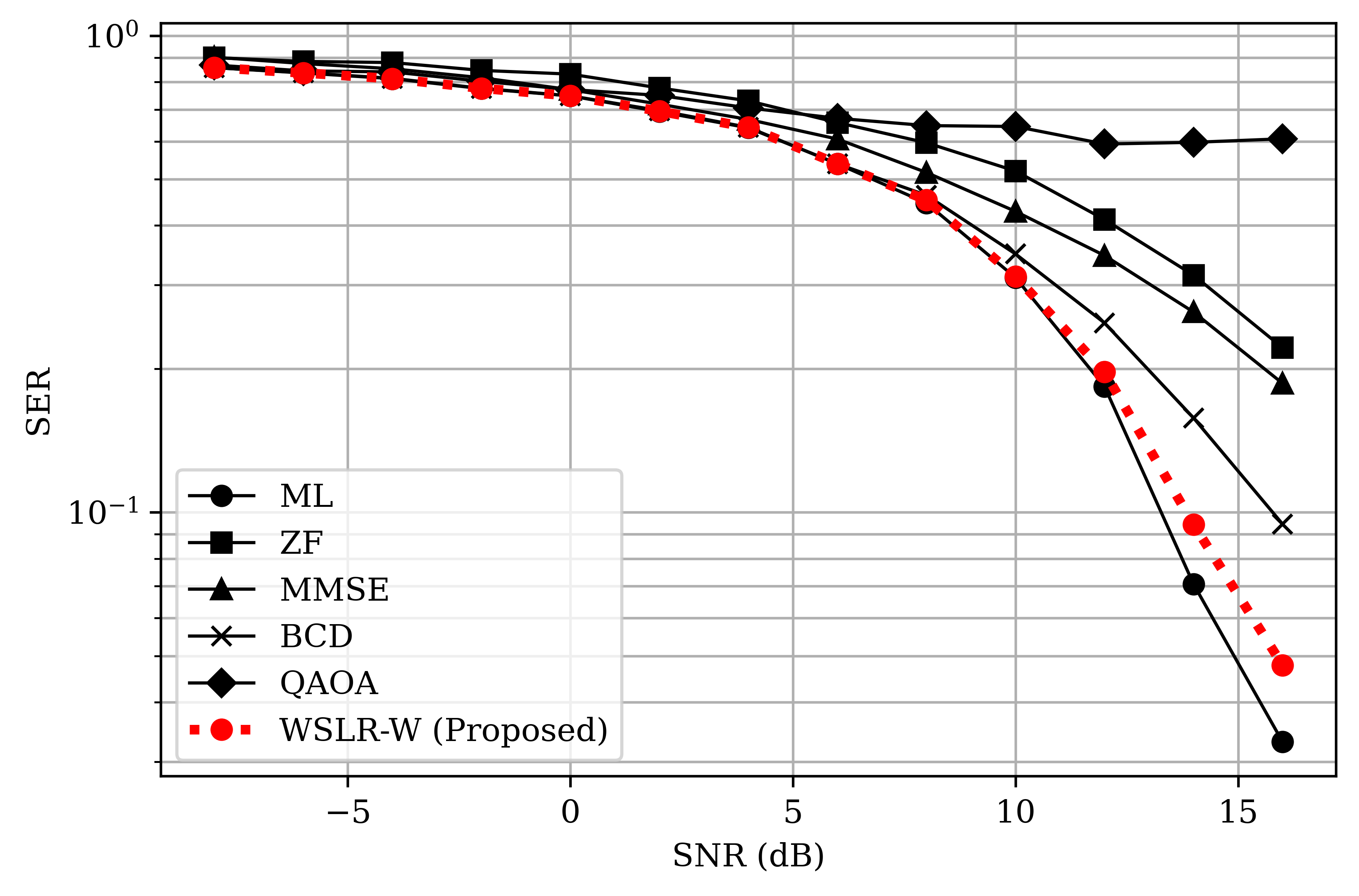}

    }

    \caption{Comparison of SER vs SNR performance with $16$-QAM modulation for different MIMO detection techniques. The quantum algorithms consider only shot noise.}
    \label{fig:overall_comparison}
\end{figure*}

\begin{table}\label{table2}
\centering
\caption{SER vs SNR comparison with realistic quantum Hardware noise models (IBM Quantum Eagle r3 processor) for $2 \times 2$ MIMO for $16$-QAM }
\label{PERCLVSQAOA}
\resizebox{0.5\textwidth}{!}{
\begin{tabular}{|c| c|c| c |c| c |c| c| c|c|}

\hline
\textbf{SNR} & \textbf{ML}& \textbf{QAOA} & \textbf{WS-Q} & \textbf{LR-Q} & \textbf{WSLR-R} & \textbf{WSLR-W} \\
\hline

-8  & 0.8115 &  0.8320 & 0.8130 & 0.8350 & 0.8200 & \textbf{0.8115} \\
\hline
-3  & 0.7110 &  0.7450 & 0.7125 & 0.7920 & 0.7315 & \textbf{0.7120} \\
\hline
2  & 0.5260  & 0.6565 & 0.5305 & 0.7465 & 0.5970 & \textbf{0.5265} \\
\hline
7  & 0.2635 & 0.5600 & 0.2745 & 0.7095 & 0.4920 & \textbf{0.2710} \\
\hline
12  & 0.0735 & 0.5480 & 0.0890 & 0.6900 & 0.4285 & \textbf{0.0830} \\
\hline
17  & 0.0120 & 0.5280 & 0.0230 & 0.6890 & 0.4135 & \textbf{0.0195} \\
\hline
\end{tabular}}

\end{table}  

\begin{table}[]
    \centering
\caption{SER vs SNR comparison with real quantum hardware (IBM Fez hardware—Heron r2) for $2\times2$ MIMO with $16$-QAM }
    \label{RH}
    \begin{tabular}{|c|c|c|c|}
    \hline
        \textbf{SNR} & \textbf{ML} & \textbf{WSLR-W(Hardware)} & \textbf{WSLR-W}\\
        \hline
          -3dB &0.7150 & 0.7200& 0.7150 \\
         \hline
          2dB &0.5350 & 0.5550& 0.5450 \\
         \hline 
        7dB &0.2200 &0.3050 & 0.2650 \\
         \hline
    \end{tabular}

\end{table}

\section{Numerical Results}\label{sim}

\subsection{Simulation Setup}
We have performed simulations for $2\times2$ and $4\times4$ MIMO systems with $16$-QAM modulation using QAOA depth $p=5$ over $40{,}000$ Monte Carlo samples. The entries of the MIMO channel matrix are modeled as independent and identically distributed (i.i.d.) complex Gaussian random variables with zero mean and unit variance, i.e., $H_{i,j} \sim \mathcal{CN}(0,1)$. A new channel realization is generated for each transmitted vector. The signal-to-noise ratio (SNR) is defined as $\mathrm{SNR} = \frac{N_r E_s}{2\sigma^2}$, where $E_s=10$ is the average symbol energy of the $16$-QAM constellation, $N_0/2 = \sigma^2$ is the double-sided noise power spectral density, where $\sigma^2$ denotes the noise variance per real dimension and $N_r$ denotes the number of receive antennas\cite{4358006,1194444}. For the warm-start initialization, a temperature parameter $T=0.2$ is used, and the resulting probabilities for $x_{k,i}^*$ are clipped to the range $[0.01,\,0.99]$ to ensure numerical stability.

The parameter $\Delta$ is continuous but restricted to $\{0.25,0.5,0.75\}$ to reduce complexity. Empirically, smaller values are preferred; for example, at $\mathrm{SNR}=-3$~dB, $\Delta=0.25$ is selected in $62\%$ of trials, compared to $24\%$ for $0.5$ and $14\%$ for $0.75$. This indicates that slower (adiabatic) evolution is more robust under noise, while larger $\Delta$ promotes exploration. Since $\Delta=0.5$ offers no clear benefit, the reduced set $\{0.25,0.75\}$ captures the trade-off effectively with minimal overhead.

\subsection{Energy Landscape Topology and Parameter Space Geometry}
Figure~\ref{fig:landscape} illustrates the cost landscape $\langle H_C\rangle$ for different QAOA variants over the parameter grid $(\gamma_{\max},\beta_{\max})$ for a $2\times2$ MIMO system at $p=5$ and $\mathrm{SNR}=7$~dB. The standard QAOA and LR-QAOA exhibit highly irregular and rugged landscapes, characterized by large ranges ($181.030$, $176.170$) and high mean costs ($37.533$, $63.440$), indicating the presence of unfavorable regions that hinder efficient optimization.

Introducing warm-start initialization with the $R_X$ mixer, as in WS-QAOA (RX) and WSLR-RX, reduces both the range ($166.467$, $163.082$) and mean cost ($29.741$, $29.604$). However, the landscapes still exhibit noticeable variability. A significant improvement is observed when replacing the $R_X$ mixer with the problem-informed WS mixer. In WS-QAOA (WS) and WSLR-W, the landscape becomes substantially smoother, with reduced ranges ($56.815$, $49.637$) and significantly lower mean costs ($-34.670$, $-33.891$). In particular, the proposed WSLR-W achieves the lowest variability ($\sigma=9.632$), indicating a more stable and optimization-friendly landscape. These observations are further supported by Table~\ref{perfQAOA}, where WSLR-W achieves the lowest minimum cost ($-49.712$), the smallest range ($49.637$), and a favorable mean cost ($-33.891$), confirming its superior convergence characteristics.

\subsection{SER Performance Evaluation}

% Figure~\ref{fig:ser_all} presents the SER performance for the $2\times2$ MIMO system. As expected, the ML detector achieves the best performance, with SER decreasing from $0.8206$ at $-8$~dB to $0.0190$ at $16$~dB. Standard QAOA exhibits a poor performance than the proposed WSLR-W QAOA, while classical methods such as ZF, MMSE, and BCD perform reasonably at low SNR but degrade significantly in high-SNR regimes. In contrast, the proposed WSLR-W consistently improves upon its initialization and closely approaches ML performance across all SNR values, achieving an SER of $0.0203$ at $16$~dB without exhibiting an error floor.

Figure~\ref{fig:ser_all} presents the SER performance for the $2\times2$ MIMO system. As expected, the ML detector achieves the best performance, with the proposed WSLR-W QAOA closely approaching the ML benchmark across all SNR values, whereas the standard QAOA performs poorly. Furthermore, the classical methods such as ZF, MMSE, and BCD perform reasonably well at low SNR but degrade significantly in high-SNR regimes.

% SER decreasing from $0.8206$ at $-8$~dB to $0.0190$ at $16$~dB. Standard QAOA exhibits a poor performance than the proposed WSLR-W QAOA, while classical methods such as ZF, MMSE, and BCD perform reasonably at low SNR but degrade significantly in high-SNR regimes. In contrast, the proposed WSLR-W consistently improves upon its initialization and closely approaches ML performance across all SNR values, achieving an SER of $0.0203$ at $16$~dB without exhibiting an error floor.

Figure~\ref{fig:mimo_qaoa_4x4} evaluates the SER performance for the  $4\times4$ MIMO system. As the MIMO dimension scales from $2\times2$ to $4\times4$, the number of qubits doubles while the circuit depth remains fixed at $p=5$, resulting in an under-parameterized ansatz that is insufficient to explore the significantly enlarged search space. Consequently, the standard QAOA algorithm suffers from noticeable performance degradation and elevated error floors. Nevertheless, the proposed WSLR-W remains robust and scalable, closely tracking ML performance and achieving an SER of $0.0478$ at $16$~dB compared to the ML bound of $0.0330$.
% \textbf{\color{blue} Explain why the standard QAOA fails in this setting. Is it due to the small circuit depth $p=5?$}

To analyze robustness under realistic conditions, hardware noise is modeled using PennyLane's \texttt{default.mixed} device \cite{bergholm2018pennylane}, incorporating depolarizing noise channels \cite{nielsen2010quantum}:
\begin{equation}
\mathcal{E}(\rho) = (1-p)\rho + \frac{p}{3}(X\rho X + Y\rho Y + Z\rho Z).
\end{equation}
Noise parameters are derived from IBM Quantum Eagle~r3 calibration data \cite{ibmquantum2026brisbane}, including $p_{1Q}=2.639\times10^{-4}$, $p_{2Q}=8.401\times10^{-3}$, and $p_{RO}=2.76\times10^{-2}$, with coherence times $T_1=247.6\,\mu$s and $T_2=105.29\,\mu$s. Table \ref{PERCLVSQAOA} shows that even with realistic hardware noise, the proposed WSLR-W QAOA achieves an SER performance close to the ML lower bound, while significantly outperforming the other variants of the QAOA method, demonstrating its strong robustness.

Finally, the proposed method is also validated on real quantum hardware using the IBM Fez processor (Heron~r2 architecture) \cite{ibmquantum2026fez}. From the results in Table \ref{RH}, it can be observed that at $-3$~dB, the hardware result ($0.7200$) closely matches ML ($0.7150$), indicating minimal degradation in low-SNR regimes. At higher SNR values (e.g., $2$~dB and $7$~dB), performance degradation becomes more noticeable due to accumulated gate errors, decoherence, and readout noise. Despite these effects, the results remain competitive, confirming that WSLR-W QAOA is robust and practically implementable on current NISQ hardware for MIMO data detection.

\section{Conclusion} \label{conc}

In this paper, we have proposed a hybrid quantum-classical QAOA-based framework for MIMO detection with $M$-QAM modulation. For Gray-coded symbol mapping, the ML detection problem is formulated as an HUBO problem. 
The optimal solution is encoded into the ground state of a problem Hamiltonian, which is solved using the QAOA algorithm. We propose a warm-start QAOA algorithm in which the quantum optimization is initialized using soft estimates obtained from a classical block coordinate descent algorithm that solves the semidefinite relaxation of the ML problem. These estimates are further leveraged in the design of a problem-informed warm start QAOA mixer and combined with a hyperparameter-free linear ramp parameter strategy, forming the proposed WSLR-QAOA framework. Simulation results demonstrate that WSLR-QAOA achieves near-ML detection performance across all tested SNR values while significantly outperforming conventional linear detectors such as ZF and MMSE, and also scales more gracefully than standard QAOA as the MIMO dimension increases. Finally, validation under realistic depolarizing noise modeled from IBM Quantum Eagle calibration data and direct execution on the IBM Fez processor (Heron~r2 architecture) confirmed that the proposed framework remains robust and practically implementable on current NISQ devices. Future work will explore advanced quantum error mitigation strategies to enhance robustness on NISQ devices.

% The key contribution is a warm start QAOA algorithm where the initialization is based on the soft estimation result obtained from a classical BCD algorithm. 
% with future work targeting advanced quantum error mitigation strategies to further close the gap between simulated and hardware performance.

% \section{Acknowledgment}
% This work was supported by the University of Melbourne through the establishment of an IBM Quantum Network Hub at the University.

\begin{appendices}
\section{HUBO based QAOA formulation for 16QAM MIMO}
\subsubsection{Pauli-Z Substitution}\label{appb}

In the quantum circuit for 16-QAM, each binary variable $b_{k,i}$ is encoded into a qubit. Using the standard convention, computational basis states are mapped to the eigenvalues of the Pauli-Z operator as:
\begin{equation}
    Z_{k,i}\ket{b_{k,i}}
    = (1 - 2b_{k,i})\ket{b_{k,i}}
    = -u_{k,i}\ket{b_{k,i}}
    \label{eq:pauli_convention}
\end{equation}
Therefore $u_{k,i} = -Z_{k,i}$, and substituting it into Eq~\eqref{gray16QAM} we obtain
\begin{align}
    s_k &= u_{k,1}(2 - u_{k,2}) \notag \\
        &= (-Z_{k,1})\bigl(2 - (-Z_{k,2})\bigr) \notag \\
        &= (-Z_{k,1})(2 + Z_{k,2}) \notag \\
        &= -2Z_{k,1} - Z_{k,1}Z_{k,2}
    \label{eq:symbol_pauli}
\end{align}
% \vspace{-30pt}

\subsubsection{HUBO Hamiltonian Construction}\label{APPA}  
The HUBO Hamiltonian can be simplified by substituting the Pauli-Z expression~\eqref{eq:symbol_pauli}, into each term
of~\eqref{ob}.
% \begin{equation*}
% H_C = \sum_{l=1}^{2N_t}\sum_{k=1}^{2N_t} G_{l,k}s_ls_k
% -\sum_{k=1}^{2N_t} 2c_k s_k\;.
% \end{equation*}
% \subsubsection*{Linear Term}
The linear term in the Hamiltonian can be simplified as
% Substituting~\eqref{eq:symbol_pauli} into $-2\sum_{k}c_k s_k$:
% %
\begin{align}
    -2\sum_{k=1}^{2N_t} c_k s_k
    &= -2\sum_{k=1}^{2N_t} c_k
       \bigl(-2Z_{k,1} - Z_{k,1}Z_{k,2}\bigr)
    \notag \\
    &= 4\sum_{k=1}^{2N_t} c_k Z_{k,1}
     + 2\sum_{k=1}^{2N_t} c_k Z_{k,1}Z_{k,2}
    \label{eq:linear_terms}
\end{align}
% \subsubsection*{Quadratic Term -- }
Next, the off-diagonal ($l \neq k$) entries of the quadratic term can be simplified as
% For $l \neq k$, substituting~\eqref{eq:symbol_pauli} into $s_l s_k$:
\begin{align}
    s_l s_k
    &= \bigl(-2Z_{l,1} - Z_{l,1}Z_{l,2}\bigr)
       \bigl(-2Z_{k,1} - Z_{k,1}Z_{k,2}\bigr)
    \notag \\
    &= 4Z_{l,1}Z_{k,1}
     + 2Z_{l,1}Z_{k,1}Z_{k,2}
     + 2Z_{l,1}Z_{l,2}Z_{k,1} \notag \\
     &+ Z_{l,1}Z_{l,2}Z_{k,1}Z_{k,2}
    \label{eq:quad_expansion}
\end{align}

Since $\mathbf{G} = \boldsymbol{\mathcal{M}}^\top\boldsymbol{\mathcal{M}}$ is symmetric,
$G_{l,k} = G_{k,l}$, so summing over all $l \neq k$ and collecting
the symmetrized pairs $l < k$ introduces a factor of 2 for each
cross term:
\begin{align}
    \sum_{\substack{l,k=1 \\ l \neq k}}^{2N_t}
    G_{l,k}\,s_l s_k
    &= 8 \sum_{l<k} G_{l,k}\,Z_{l,1}Z_{k,1}
     + 4 \sum_{l<k} G_{l,k}\,Z_{l,1}Z_{k,1}Z_{k,2}
    \nonumber \\
    &+ 4 \sum_{l<k} G_{l,k}\,Z_{l,1}Z_{l,2}Z_{k,1}
    \nonumber  \\
     &+ 2 \sum_{l<k} G_{l,k}\,Z_{l,1}Z_{l,2}Z_{k,1}Z_{k,2}
     \label{eq:off_diag}
\end{align}
% where each coefficient arises from $2 \times$ the corresponding
% coefficient in~\eqref{eq:quad_expansion}:
% $2\times4=8$, $2\times2=4$, $2\times2=4$, $2\times1=2$.

% \subsubsection*{Quadratic Term -- Diagonal ($l = k$)}

The diagonal entries $l = k$ of the quadratic term can be simplified by using the idempotent property of Pauli operators $Z_{k,i}^2 = I$ as
\begin{align}
    s_k^2
    &= \bigl(-2Z_{k,1} - Z_{k,1}Z_{k,2}\bigr)^2
    \nonumber \\
    &= 4Z_{k,1}^2
     + 2\cdot(-2Z_{k,1})(-Z_{k,1}Z_{k,2})
     + (Z_{k,1}Z_{k,2})^2
    \nonumber \\
    &= 4I + 4Z_{k,1}^2 Z_{k,2} + Z_{k,1}^2 Z_{k,2}^2
    \nonumber \\
    &= 4I + 4Z_{k,2} + I
    \nonumber \\
    &= 5I + 4Z_{k,2}
    \label{eq:diagonal_terms}
\end{align}
Summing over all diagonal entries we obtain
\begin{equation}
    \sum_{k=1}^{2N_t} G_{k,k}\,s_k^2
    = 5\sum_{k=1}^{2N_t} G_{k,k}\,I
    + 4\sum_{k=1}^{2N_t} G_{k,k}\,Z_{k,2}
    \label{eq:diag_sum}
\end{equation}
The constant term $5\sum_k G_{k,k}I$ shifts the energy spectrum
uniformly and does not affect the minimizer; therefore, it is absorbed into the identity component of the Hamiltonian and dropped.

Collecting~\eqref{eq:linear_terms}, 
and~\eqref{eq:diag_sum} into~\eqref{ob}, and discarding all
identity (constant) terms, the cost Hamiltonian expressed
entirely in Pauli-Z operators is given by Eq \eqref{16QAM_ISING}.

% \subsubsection*{Full Cost Hamiltonian}

\end{appendices}
% ================= References =================
\bibliographystyle{IEEEtran}
\bibliography{references}

\end{document}